\documentstyle[12pt,epsfig,amsfonts]{article}
\renewcommand{\theequation}{\arabic{section}.\arabic{equation}} 
\renewcommand{\thesubsection}{\arabic{section}.\arabic{subsection}.}

\title{Radiation reaction, renormalization\\ and conservation laws\\ in 
six-dimensional classical electrodynamics}
\author{\bf Yurij Yaremko\footnote{Electronic mail: yar@ph.icmp.lviv.ua}}
\date{\it Institute for Condensed Matter Physics, \\
1 Svientsitskii St., 79011 Lviv, Ukraine}
\begin{document}

\maketitle

\begin{abstract}
A self-action problem for a point-like charged particle arbitrarily moving 
in flat space-time of six dimensions is considered. A consistent 
regularization procedure is proposed which relies on energy-momentum and 
angular momentum balance equations. Structure of the angular momentum tensor 
carried by the retarded "Li\'enard-Wiechert" field testifies that a 
point-like source in six dimensions possesses an internal angular 
momentum. Its magnitude is proportional to the square of acceleration. It is 
the so-called {\em rigid} relativistic particle; its motion is determinated 
by the higher-derivative Lagrangian depending on the curvature of the world 
line. It is shown that action functional contains, apart from usual "bare" 
mass, an additional renormalization constant which corresponds to the 
magnitude of "bare" internal angular momentum of the particle.
\end{abstract}
PACS numbers: 03.50.De, 11.10.Gh

\section{Introduction}

Recently \cite{Gl,KLS}, there has been considerable interest in 
renormalization procedure in classical electrodynamics of a point particle 
moving in flat space-time of arbitrary dimensions. The main task is to 
derive the analogue of the well-known Lorentz-Dirac equation \cite{Dir}. 
The Lorentz-Dirac equation is an equation of motion for a charged particle 
under the influence of an external force as well as its own electromagnetic 
field. (For a modern review see \cite{Rohr,Pois,TVW}.) In earlier paper 
\cite{Kos} the Lorentz-Dirac equation in six dimensions is obtained via the 
consideration of energy-momentum conservation.

All the authors \cite{Gl,KLS,Kos} deal with an obvious generalization of the 
standard variational principle used in four dimensions
\begin{equation}\label{I}
I = I_{\mbox{\scriptsize particle}} + 
I_{\mbox{\scriptsize int}} + I_{\mbox{\scriptsize field}}\,,
\end{equation}
with
\begin{equation} \label{Mp}
I_{\mbox{\scriptsize 
field}}=-\frac{1}{4\Omega_{D-2}}\int {\rm d}^DyF^{\mu\nu}F_{\mu\nu}\,,
\quad
I_{\mbox{\scriptsize particle}}=-m\int {\rm d}\tau\sqrt{-{\dot z}^2}\,,
\end{equation}
and the interaction term given by
\begin{equation}
I_{\mbox{\scriptsize int}} = e\int {\rm d}\tau A_\mu 
{\dot z}{}^\mu .
\end{equation}
By $\Omega_{D-2}$ the area of a $(D-2)$-dimensional sphere of unit 
radius is denoted:
\begin{equation} \label{OD}
\Omega_{D-2}=2\frac{\pi^{(D-1)/2}}{\Gamma(\frac{D-1}{2})}.
\end{equation}

Strictly speaking, the action integral (\ref{I}) may be used to derive 
trajectories of the test particles, when the field is given {\em a 
priory}. It may also be used to derive $D-$dimensional Maxwell equations, 
if the particle trajectories are given {\em a priory}. Simultaneous 
variation with respect to both field and particle variables is incompatible
since the Lorentz force will always be ill defined in the immediate 
vicinity of the particle's world line. 

The elimination of the divergent self-energy of a point charge is the key 
to the problem. In four-dimensional space-time one usually assumes that the 
parameter $m$ involving in $I_{\mbox{\scriptsize particle}}$ is the 
unphysical {\em bare} mass. It absorbs the inevitable infinity within the 
renormalization procedure and becomes the {\em observable} rest mass of the 
particle. In $D$ dimensions the Coulomb potential of a charge scales as 
$|{\bf x}|^{3-D}$ \cite{KosPr}. Inevitable infinities arising in 
higher-dimensional electrodynamics are stronger than in four dimensions. 

All the authors \cite{Gl,KLS,Kos} agree, that in even dimensions higher than 
four divergences cannot be removed by the renormalization of mass included 
in the initial action integral (\ref{I}). To make classical electrodynamics in six dimensions a renormalizable theory,
in \cite{Kos} the six-dimensional analogue of the relativistic particle 
with rigidity \cite{Pl,Pv,Nest} is substituted for the structureless point 
charge whose action term is proportional to worldline length. Corresponding 
Lagrangian involves, apart from usual "bare mass", an additional 
regularization constant which absorbs one extra divergent term. In 
\cite{KLS} the procedure of regularization in any dimensions is elaborated. 
It allows to remove the infinities coming from the particle's self-action by 
introducing new counterterms in the particle action. 

On the contrary, Gal'tsov \cite{Gl} states that the theory is 
nonrenormalizable in dimensions higher than four. Introduction of higher 
derivatives in the particle action term seems for him not reasonable 
enough. 

In the present paper the problem of renormalizability will be reformulated 
within the problem of Poincar\'e invariance of a closed particle plus field 
system. The conservation laws are an immovable fulcrum about which tips the 
balance of truth regarding renormalization and radiation reaction. Either 
nonrenormalizable theory or renormalizable one should be compatible with the 
Poincar\'e symmetry.

In \cite{Teit} a strict geometrical sense of divergent terms arising in four 
dimensions is made. Teitelboim calculates how many energy-momentum of the 
retarded electromagnetic field of an arbitrarily moving point charge flows 
across a thin tube around the world line. The energy-momentum contains two 
quite different terms: (i) the bound part which is permanently "attached" to 
the charge and is carried along with it; (ii) the radiation part detaches 
itself from the charge and leads an independent existence. The former is 
divergent while the latter is finite (the integral of the Larmor 
relativistic rate of radiated energy-momentum over particle's world line 
is meant). Hence, a charged particle can not be separated from its bound 
electromagnetic "cloud", so that the four-momentum of the particle is the 
sum of the mechanical momentum and the electromagnetic bound four-momentum.
The electromagnetic part contains, apart from divergent self-energy which is 
linked with the bare mass, also a finite structureless term which is 
proportional to  the particle acceleration. On rearrangement, Teitelboim's 
expression for the four-momentum of accelerated point-like charge looks as 
follows 
\begin{equation} \label{Texp}
p_{\mbox{\scriptsize part}}^\mu = mu^\mu - \frac{2}{3}e^2a^\mu\,,
\end{equation}
where $m$ is already renormalized rest mass.

Similar decomposition of the angular momentum tensor of the retarded 
Li\'enard-Wiechert field into the bound and the radiative components is 
performed in \cite{LV}. It is shown, that the bound electromagnetic "cloud" 
possesses its own angular momentum. It has precisely the same form as the 
mechanical angular momentum of a "bare" charge. Therefore, the 
regularization of angular momentum can be reduced to the renormalization of 
mass.

Conserved quantities place stringent requirements on the dynamics of the 
system. They demand that the change in {\em radiative} energy-momentum and 
angular momentum should be balanced by a corresponding change in the 
{\em already renormalized} momentum and angular momentum of the particle. It 
is shown \cite{Yar} that energy-momentum balance equation gives the 
relativistic generalization of Newton's second law where loss of energy due 
to radiation is taken into account. The angular momentum balance equation 
explains how four-momentum of charged particle depends on its velocity and 
acceleration (see eq.(\ref{Texp})). So, a careful analysis with the use of 
regularization procedure compatible with the Poincar\'e symmetry leads 
to the Lorentz-Dirac equation in four-dimensional case.

In this paper we calculate the energy-momentum and angular momentum of the 
retarded electromagnetic field generated by a point-like charge in six 
dimensions. The form of (divergent) bound parts reveals the structure of a 
"bare" core "dressed" in the electromagnetic cloud. Does a 
consistent classical electrodynamics in spacetimes of dimensions 
$D>4$ lead inevitably to the rigid particle? If so, that the bound 
characteristics possess the specific features, e.g. the internal angular 
momentum.

Further we analyse the radiative parts. We see that the energy-momentum and 
angular momentum balance equations allow us to establish the radiation 
reaction force in four dimensions \cite{Yar}. Does a consistent classical 
electrodynamics in spacetimes of dimensions $D>4$ lead inevitably to the 
rigid particle? If so, that the Poincar\'e conservation laws give the 
corresponding radiation reaction force.

\section{General setting}
\setcounter{equation}{0} 
Let ${\mathbb M}_6$ be 6-dimensional Minkowski space with coordinates
$y^\mu$ and metric tensor $\eta_{\mu\nu}={\mbox{\rm diag}}(-1,1,1,1,1,1)$.
We use the natural system of units with the velocity of light $c=1$. 
Summation over repeated indices is understood throughout the paper; Greek 
indices run from $0$ to $5$, and Latin indices  from $1$ to $5$.

We consider an electromagnetic field $F_{\alpha\beta}$ produced by a 
point-like particle of charge $e$. The particle moves in flat space-time 
${\mathbb M}_6$ on an arbitrary world line 
\begin{eqnarray} \label{trj}
\zeta&:&{\mathbb R}\to {\mathbb M}_6\nonumber\\
&&u\mapsto (z^\mu(u))\,,
\end{eqnarray}
where $u$ is proper time. The Maxwell field equation are
\begin{equation}
F^{\alpha\beta}{}_{,\,\beta}=\frac{8\pi^2}{3}j^\alpha
\end{equation}
where current density $j^\alpha$ is given by
\begin{equation}\label{j}
j^\alpha =e\int {\rm d}\tau u^\alpha(u)\delta(y-z(u)).
\end{equation}
$u^\alpha (u)$ denotes the (normalized) six-velocity vector 
${\rm d} z^\alpha(u)/{\rm d} u$ and the factor $8\pi^2/3$ is the area of 
4-dimensional unit sphere embedded in ${\mathbb M}_6$ (see eq.(\ref{OD}) for
$D=6$).

We express the electromagnetic field in terms of a vector potential, 
$\hat F={\rm d}\hat A$. In the Lorentz gauge $A^\alpha{}_{\,,\alpha}=0$ 
the Maxwell field equations become
\begin{equation}
\square A^\alpha(y)=-\frac{8\pi^2}{3}j^\alpha(y)
\end{equation}
where $\square:=\eta^{\alpha\beta}\partial_\alpha\partial_\beta$ is the 
wave operator. Using the retarded Green function \cite[eq.(3.4)]{Gl} 
associated with the D'Alembert operator $\square$ and the charge-current 
density vector (\ref{j}) we construct the retarded Li\'enard-Wiechert 
potential in six dimensions:
\begin{equation}\label{A}
A_\mu(y)=e\int {\rm d} u u_\mu(u)\left(-\frac{1}{2\pi 
R}\frac{{\rm d}}{{\rm d} R}\frac{\delta(T-R)}{R}\right).
\end{equation}
Here $T:=y^0-z^0(u)$ and $R:=|{\bf y}-{\bf z}(u)|$.

We suppose that the dynamics of our composite particle plus field system is 
governed by the conservation laws which arise from the invariance of the 
closed system under time and space translations as well as space and mixed 
space-time rotations. The components of momentum 6-vector carried by the 
electromagnetic field are \cite{Kos}
\begin{equation} \label{pem}
p_{\mbox{\scriptsize em}}^\nu (\tau)=P\int_{\Sigma}
{\rm d}\sigma_\mu T^{\mu\nu} 
\end{equation}
where ${\rm d}\sigma_\mu$ is the vectorial surface element on an arbitrary 
space-like hypersurface $\Sigma$. The components of the electromagnetic 
field's stress-energy tensor 
\begin{equation}\label{T}
\frac{8\pi^2}{3}T^{\mu\nu} = F^{\mu\lambda}F^\nu{}_\lambda - 
1/4\eta^{\mu\nu} F^{\kappa\lambda}F_{\kappa\lambda}
\end{equation}
have singularities on a particle trajectory (\ref{trj}). In 
eq.(\ref{pem}) capital letter $P$ denotes the principal value of the 
singular integral, defined by removing from $\Sigma$ an 
$\varepsilon$-sphere around the particle and then passing to the limit 
$\varepsilon\to 0$.

The angular momentum tensor of the electromagnetic field is written as
\cite{Rohr}
\begin{equation} \label{Mem}
M_{\mbox{\scriptsize em}}^{\mu\nu}(\tau)=P\int_{\Sigma}
{\rm d}\sigma_\alpha\left(y^\mu T^{\alpha\nu}-y^\nu T^{\alpha\mu}\right) .
\end{equation}

\begin{figure}
\begin{center}
\epsfclipon
\epsfig{file=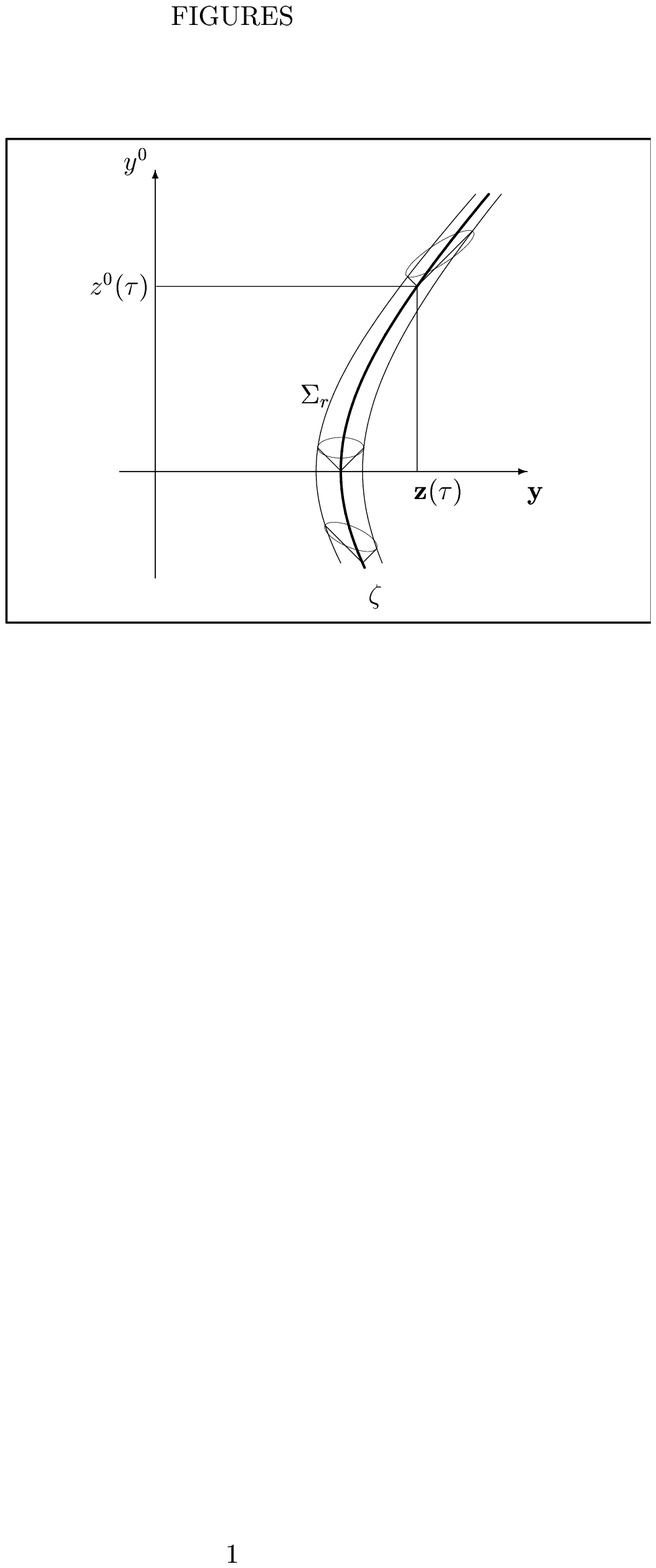,width=8cm}
\end{center}
\caption{\label{tube}
Integration region considered in the evaluation of the bound and 
emitted conserved quantities produced by all points of the world line up to 
the end point {\sf E}. Retarded spheres $S(z(u),r), u\in]-\infty,\tau],$ 
of constant radii $r$ constitute a thin world tube $\Sigma_r$ enclosing the 
world line $\zeta$. The sphere $S(z(u),r)$ is the intersection of the future 
light cone with vertex at point $z^\mu(u)\in\zeta$ and $r-$shifted hyperplane
$\Sigma(z(u),r)$ which is orthogonal to six-velocity $u^\mu(u)$.}
\end{figure}

Kosyakov \cite{Kos} calculates the radiative part of the energy-momentum 
(\ref{pem}) which flows across a world tube of constant radius $r$ enclosing 
the world line $\zeta$. This integration hypersurface, say $\Sigma_r$, is a 
disjoint union of (retarded) spheres of constant radii $r$ centered on a 
world line of the particle (see Fig.\ref{tube}). The sphere $S(z(u),r)$ is 
the intersection of future light cone generated by null rays emanating from 
$z(u)\in\zeta$ in all possible directions
\begin{equation}\label{C}
C(z(u))=\{y\in {\mathbb M}_6: 
(y^0-z^0(u))^2=\sum_i(y^i-z^i(u))^2,y^0-z^0(u)>0\},
\end{equation}
and tilted hyperplane
\begin{equation} \label{Ksi}
\Sigma(z(u),r)=\{y\in {\mathbb M}_6: 
u_\alpha(u)(y^\alpha -z^\alpha(u)-u^\alpha(u)r)=0\}.
\end{equation}

\begin{figure}
\begin{center}
\epsfclipon
\epsfig{file=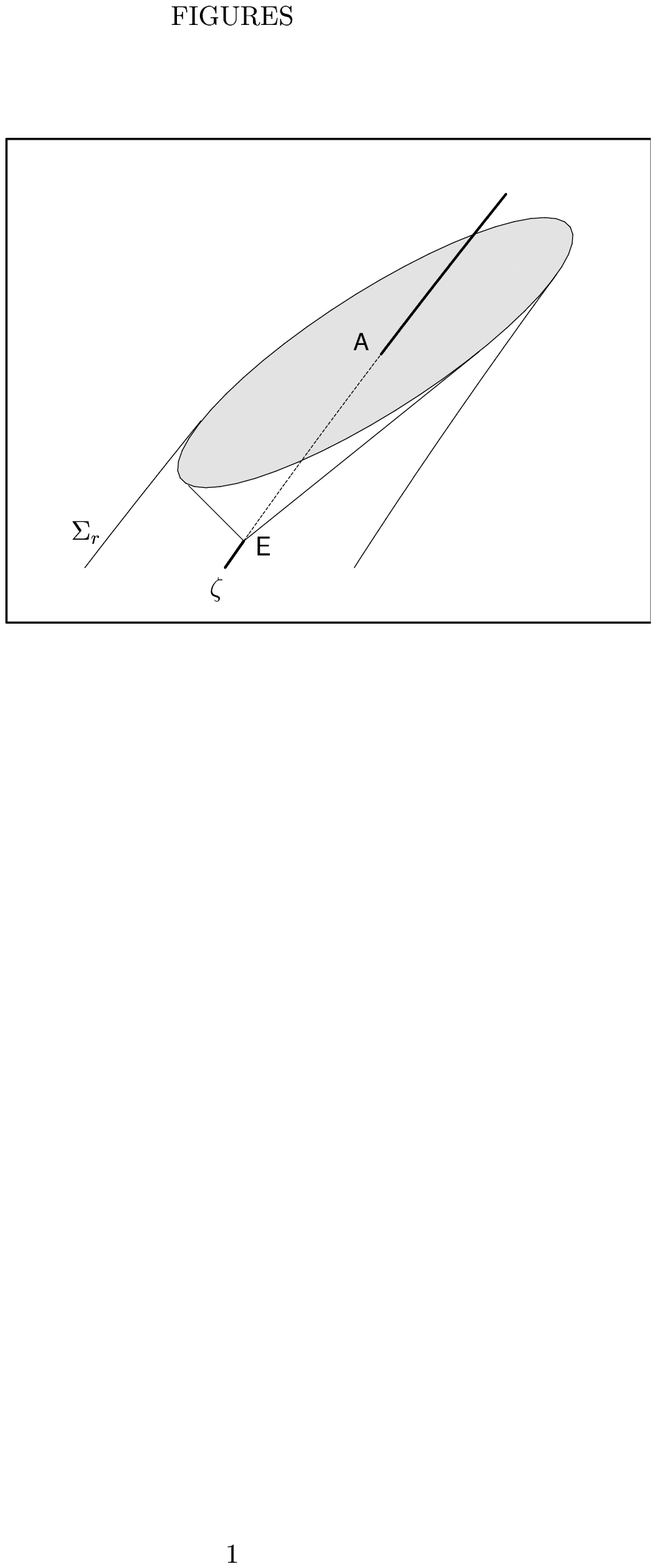,width=9cm}
\end{center}
\caption{\label{cap}
Integration region considered in the evaluation of the 
energy-momentum and angular momentum produced by the upper segment {\sf EA}
of the world line. Corresponding electromagnetic field conserved quantities 
flow across the cap covering the top of world tube $\Sigma_r$. The cap is 
the part of the observation hyperplane $\Sigma(z(\tau),r)$. Point {\sf E}
has coordinates $(z^\mu(\tau))$ and point {\sf 
A}$=\zeta\cup\Sigma(z(\tau),r)$.
}
\end{figure}

Integration surface $\Sigma_r$ is time-like. For the sake of completeness, 
the narrow cylindrical tube surrounding the world line should be 
covered by a space-like "cap" (see Fig.\ref{cap}). To evaluate the part of 
energy-momentum and angular momentum prodused by the upper segment {\sf EA} 
of the world line, we integrate (\ref{pem}) and (\ref{Mem}) over space-like 
hyperplane $\Sigma(z(\tau),r)$. For the sake of simplicity we make such a 
Lorentz transformation that this tilted hyperplane becomes 
$\Sigma_{t'}=\{y\in{\mathbb M}_6:y^{0'}=t'\}$. The Lorentz matrix, 
$\Lambda(\tau)$, determines the transformation to the particle's momentarily 
comoving Lorentz frame (MCLF) where the particle is momentarily at rest at 
observation instant $\tau$. On rearrangement, energy-momentum (\ref{pem}) 
and angular momentum (\ref{Mem}) take the form
\begin{eqnarray}\label{p}
p_{\mbox{\scriptsize em}}^\nu 
(\tau)&=&\Lambda^\nu{}_{\nu'}(\tau)P\int_{\Sigma_{t'}}
{\rm d}\sigma_{0'} T^{0'\nu'} \\ 
M_{\mbox{\scriptsize em}}^{\mu\nu}(\tau)&=&\Lambda^\mu{}_{\mu'}(\tau)
\Lambda^\nu{}_{\nu'}(\tau)P\int_{\Sigma_{t'}}
{\rm d}\sigma_{0'}\left(y^{\mu'}T^{0'\nu'}-y^{\nu'}T^{0'\mu'}\right). 
\label{M}
\end{eqnarray}

\section{Coordinate systems}
\begin{figure}
\begin{center}
\epsfclipon
\epsfig{file=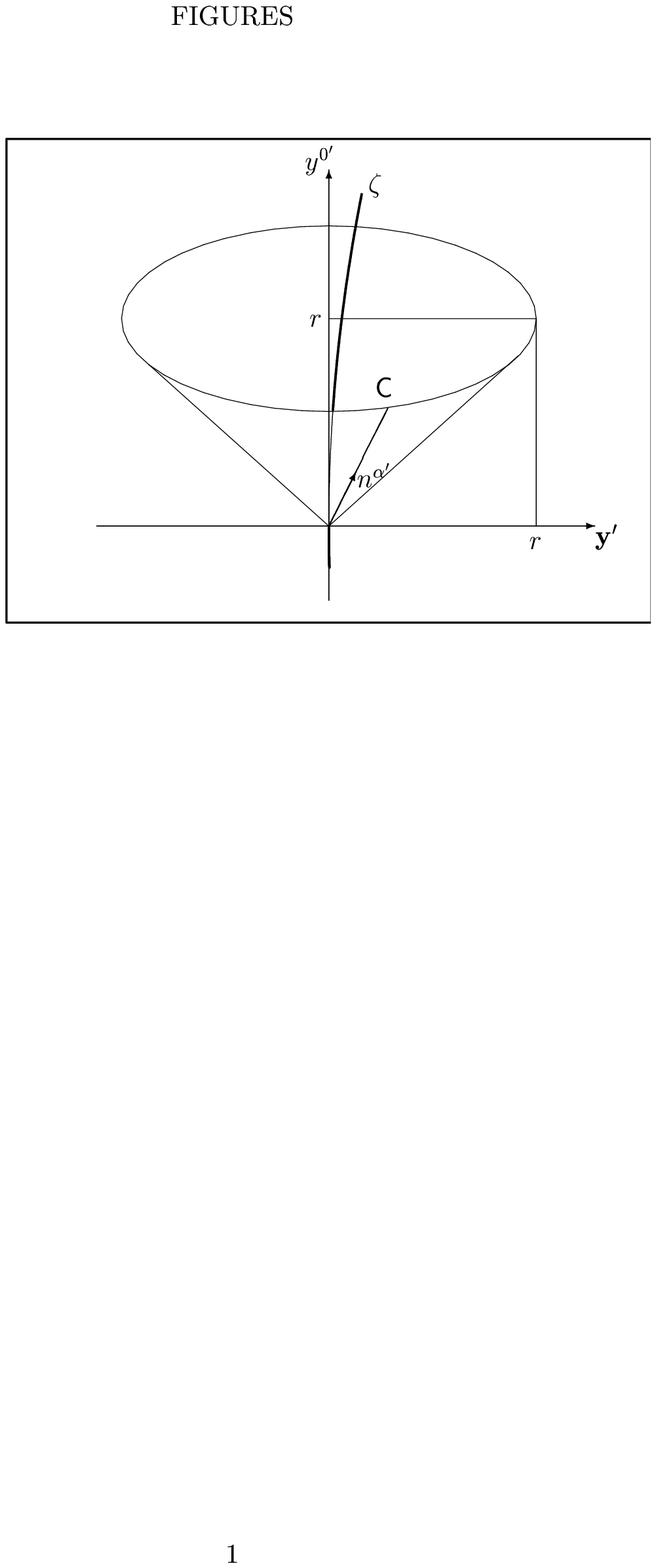,width=10cm}
\end{center}
\caption{\label{cone}
In MCLF the retarded distance is the distance between any point on 
spherical light front $S(0,r)=\{y\in {\mathbb M}_6: 
(y^{0'})^2=\sum_i(y^{i'})^2,y^{0'}=r>0\}$ and the particle. The charge is 
placed in the coordinate origin; it is momentarily at the rest.
The point {\sf C}$\in S(0,r)$ is linked to the coordinate origin by a null 
ray characterized by the angles $\vartheta^A$ specifying its direction on 
the cone. The vector with components $n^{\alpha'}$ is tangent to this null 
ray.
}
\end{figure}

\setcounter{equation}{0} 
An appropriate coordinate system is a very important for the integration. 
We use an obvious generalization of {\em a coordinate system 
centered on an accelerated world line}\cite{NU,Pois}. The set of curvilinear 
coordinates for flat spacetime ${\mathbb M}_6$ involves the {\it retarded 
time} $u$ and the {\it retarded distance} $r$ introduced in previous 
section (see eqs.(\ref{C}) (\ref{Ksi})). To understand the situation more 
thoroughly, we pass to the MCLF where the particle is momentarily at the 
rest at the retarded time $u$. The retarded distance $r$ is the distance 
between an observer event {\sf C}$\in\Sigma_r$ and the particle, as measured 
at $u$ in the MCLF (see Fig.\ref{cone}). Points on the sphere 
$S(0,r)\subset\Sigma_r$ are distinguished by four spherical angles 
$(\phi,\vartheta_1,\vartheta_2,\vartheta_3)$:
\begin{eqnarray} \label{fit}
y^{0'}=r,\qquad 
&&y^{1'}=r\cos\phi\sin\vartheta_1\sin\vartheta_2\sin\vartheta_3
\nonumber\\
&&y^{2'}=r\sin\phi\sin\vartheta_1\sin\vartheta_2\sin\vartheta_3
\nonumber\\
&&y^{3'}=r\cos\vartheta_1\sin\vartheta_2\sin\vartheta_3
\nonumber\\
&&y^{4'}=r\cos\vartheta_2\sin\vartheta_3
\nonumber\\
&&y^{5'}=r\cos\vartheta_3 
\end{eqnarray} 
In the laboratory frame the points on this sphere have the following 
coordinates:
\begin{eqnarray}\label{y_r}
y^\alpha&=&z^\alpha (u) + r\Lambda^\alpha{}_{\alpha'}(u)n^{\alpha'}
\nonumber\\
&=&z^\alpha (u) + rk^\alpha
\end{eqnarray}
where $n^{\alpha'}=(1,n^{i'})$ is null vector with space components given by 
eqs.(\ref{fit}).

We see that flat space-time ${\mathbb M}_6$ becomes a disjoint union of 
world tubes $\Sigma_r, r>0,$ enclosing the particle trajectory (\ref{trj}). 
A world tube  is a disjoint union of (retarded) spheres of constant radii 
$r$ centered on a world line of the particle. Points on a sphere are 
distinguished by four spherical polar angles.

If we take the integration hypersurface $\Sigma_r$ being a surface of 
constant $r$, these parametrization of ${\mathbb M}_6$ is suitable. If we 
integrate over hyperplane $\Sigma_t=\{y\in{\mathbb M}_6:y^0=t\}$, we need
another curvilinear coordinates. We substitute the expression 
$(t-u)/k^0$ for the retarded distance $r$ in eqs.(\ref{y_r}). Particle 
trajectory now is meant as a local section 
\begin{eqnarray} 
\label{tri}
\zeta &:&{\mathbb R}\to {\mathbb M}_6 \nonumber\\
&&u\mapsto (u,z^i(u))
\end{eqnarray}
of trivial bundle $({\mathbb M}_6,i,{\mathbb R})$ where the projection 
\cite{Snd}
\begin{eqnarray} \label{i}
i&:&{\mathbb M}_6\to {\mathbb R} \nonumber\\
&&(y^0,y^i)\mapsto y^0
\end{eqnarray} 
defines the instant form of dynamics \cite{GKT}. The final coordinate 
transformation $(y^\alpha)\mapsto 
(t,u,\vartheta_1,\vartheta_2,\vartheta_3,\phi)$ looks as follows:
\begin{equation} \label{flat}
y^0=t, \qquad y^i=z^i(u) + 
(t-u)\frac{k^i}{k^0}.
\end{equation}
The integration hyperplane $\Sigma_t$ is a surface of constant $t$.

Flat space-time ${\mathbb M}_6$ becomes a disjoint union of 
fibres $i^{-1}(t):=\Sigma_t$ of the trivial bundle (\ref{i}). A fibre 
$\Sigma_t$ is a disjoint union of (retarded) spheres
centered on a world line of the particle. The sphere 
\[
S(z(u),t-u)=\{y\in 
{\mathbb M}_6: (y^0-u)^2=\sum_i(y^i-z^i(u))^2,y^0=t,t-u>0\}
\]
is the intersection of future light cone, generated by null rays emanating 
from $z(u)\in\zeta$ in all possible directions, and hyperplane $\Sigma_t$.
For the fixed instant $t$ the retarded time parameter $u\in 
]-\infty,t]\subset\mathbb R$. Points on the sphere are distinguished by 
spherical polar angles (see eqs.(\ref{fit})).

\section{Electromagnetic potential and electromagnetic field in six 
dimensions}
\setcounter{equation}{0} 
In even space-time dimensions the Green function associated with the 
D'Alembert operator is localized on the light cone \cite{Gl,KLS}. Having 
integrated (\ref{A}) we obtain
\begin{eqnarray} \label{Am}
 A_\mu&=&\frac{e}{2\pi}\frac{1}{r}\frac{{\rm d}}{{\rm d} 
u}\left(\frac{u_\mu(u)}{r}\right)
\nonumber\\
&=&\frac{e}{2\pi}\left[\frac{a_\mu(u)}{r^2}+
\frac{u_\mu(u)}{r^3}\left(1+ra_k\right)\right]
\end{eqnarray}
where $a_k=a_\alpha k^\alpha$ is the component of the acceleration 
$a_\alpha={\rm d} u_\alpha/{\rm d} u$ in the direction of $k^\alpha$. It 
is understood that in eq.(\ref{Am}), all world line quantities (such as 
$u_\mu$ and $a_\mu$) are to be evaluated at the retarded time $u$.

The potential (\ref{Am}) differs from that of \cite[eq.(13)]{Kos} just by an 
overall coefficient $e/2\pi$.

The direct particle field \cite{HN} is defined in terms of this potential by
$F_{\alpha\beta}=A_{\beta ,\alpha}-A_{\alpha ,\beta}$. Having used the 
differentiation rule \cite[eqs.(2),(3)]{Kos}
\begin{equation}
\frac{\partial u}{\partial x^\mu}=-k_\mu,\qquad  
\frac{\partial r}{\partial x^\mu}=-u_\mu+\left(1+ra_k\right)k_\mu,
\end{equation}  
we obtain
\begin{equation} \label{F}
F=\frac{e}{2\pi}\left(\frac{u\wedge a}{r^3}+V\wedge k\right)
\end{equation}
where
\begin{equation} \label{V}
V_\mu=\frac{3u_\mu}{r^4}+\frac{3(a_\mu +2u_\mu a_k)}{r^3}+\frac{{\dot a}_\mu 
+u_\mu {\dot a}_k + 3a_\mu a_k +3u_\mu a^2_k}{r^2}.
\end{equation}
The overdot means the derivative with respect to retarded time $u$.
Li\'enard-Wiechert field (\ref{F}) coincides with the field obtained in 
\cite[eq.(14)]{Kos} where the "mostly minus" metric signature should be 
replaced by the "mostly plus" one.

\section{Energy-momentum of the retarded Li\'enard-Wi\-e\-chert field in 
six dimensions}
\setcounter{equation}{0} 

It is straightforward to substitute the components (\ref{F}) into 
eq.(\ref{T}) to calculate the electromagnetic field's stress-energy tensor. 
Following \cite{Kos}, we present $T^{\alpha\beta}$ as a sum of 
radiative and bound components,
\begin{equation} \label{Tr_Tb}
T^{\alpha\beta} = T_{\mbox{\scriptsize rad}}^{\alpha\beta} + 
T_{\mbox{\scriptsize bnd}}^{\alpha\beta}.
\end{equation}
The radiative part scales as $r^{-4}$:
\begin{equation} \label{Tr} 
\frac{8\pi^2}{3}T_{\mbox{\scriptsize rad}}^{\alpha\beta}=
\frac{e^2}{4\pi^2}\frac{k^\alpha k^\beta}{r^4}V_{(-2)}^\mu 
V_\mu^{(-2)}
\end{equation}
where the components $V_\mu^{(-2)}$ of six-vector $V_{(-2)}$ is defined by
eq.(\ref{V}). The others $T_{(-\kappa)}$ constitute the bound part of the 
Maxwell energy-momentum tensor density:
\begin{equation}
T_{\mbox{\scriptsize bnd}}^{\alpha\beta}=T_{(-8)}+T_{(-7)}+T_{(-6)}+T_{(-5)}.
\end{equation}
(Each term has been labelled according to its dependence on the distance $r$.)

According \cite{Kos}, the outward-directed surface element ${\rm 
d}\sigma_\mu$ of a five-cylinder $r=const$ in ${\mathbb M}_6$ is
\begin{equation} \label{sgm} 
{\rm d}\sigma_\mu=\left[-u_\mu+(1+ra_k)k_\mu\right]r^4{\rm d}\Omega_4{\rm 
d} u
\end{equation}
where 
${\rm 
d}\Omega_4={\rm 
d}\vartheta_1{\rm d}\vartheta_2{\rm d}\vartheta_3{\rm d}\phi\sin\vartheta_1
\sin^2\vartheta_2\sin^3\vartheta_3$ is the element of solid angle in five
dimensions. The angular integration can be handled via the relations
\begin{eqnarray}
\int {\rm d}\Omega_4=\frac{8\pi^2}{3},\qquad
\int {\rm d}\Omega_4n^\alpha n^\beta=\frac{8\pi^2}{15}\left(
\eta^{\alpha\beta}+u^\alpha u^\beta\right),\nonumber\\
\int {\rm d}\Omega_4n^\alpha n^\beta n^\gamma n^\kappa 
=\frac{8\pi^2}{105}\left[
\left(\eta^{\alpha\beta}+u^\alpha u^\beta\right)
\left(\eta^{\gamma\kappa}+u^\gamma u^\kappa\right) 
 \right.
\nonumber\\
+\left.\left(\eta^{\alpha\gamma}+u^\alpha u^\gamma\right)
\left(\eta^{\beta\kappa}+u^\beta 
u^\kappa\right) +
\left(\eta^{\alpha\kappa}+u^\alpha u^\kappa\right)
\left(\eta^{\beta\gamma}+u^\beta u^\gamma\right) \right]
\end{eqnarray}
The integral of polynomial in odd powers of $n^\alpha:=k^\alpha 
-u^\alpha$ vanishes.

We are now concerned with volume integration of (\ref{pem}). Although the    
surface element (\ref{sgm}) contains the term which is proportional to 
$r$, the radiative part of electromagnetic-field six-momentum 
$p_{\mbox{\scriptsize rad}}$ does not depend on the distance:
\begin{equation} \label{prad}
p_{\mbox{\scriptsize rad}}^\mu=\frac{e^2}{4\pi^2}
\int_{-\infty}^\tau {\rm d} u
\left(\frac45u^\mu\dot{a}^2-\frac{6}{35}a^2\dot{a}^\mu
+\frac37a^\mu (a^2)^\cdot+2a^4u^\mu\right)
\end{equation}
(We denote $(a^2)^{\bf\cdot}$ the derivative ${\rm d} a^2/{\rm d} u$.) The 
reason is that $k_\alpha T_{\mbox{\scriptsize rad}}^{\alpha\beta}=0$. 
Since $k_\alpha T_{(-5)}^{\alpha\beta}=0$, this term does not produce a 
change in
radiation flux.

Volume integration of the bound part of the stress-energy tensor 
over the world tube $\Sigma_r$ of constant radius $r$ reveals 
that the bound energy-momentum is a function of the end points only:
\begin{equation} \label{pbnd}
p_{\mbox{\scriptsize bnd}}^\mu=\frac{e^2}{4\pi^2}
\left[
\frac32\frac{u^\mu(u)}{r^3}+\frac{12}{5}\frac{a^\mu(u)}{r^2} +
2\frac{a^2u^\mu(u)}{r}
\right]_{u\to -\infty}^{u=\tau}
\end{equation}
(The matter is that the total (retarded) time derivatives arise from angular 
integration.) If the charged particle is asymptotically free at the remote 
past, we obtain the Coulomb-like self-energy of constant value. The upper 
limit drastically depends on the value of $r$. If it is finite, we have no 
problem. If $r$ tends to zero, $p_{\mbox{\scriptsize bnd}}^\mu\to\infty$. A 
subtle point in this integration is that a surface of constant $r$ is 
time-like. The space-like "cap" should be added to close the time-like tube
$\Sigma_r$ (see Fig.\ref{cap}). The integration over this cap results 
$p_{\mbox{\scriptsize bnd}}^\mu$ finally.

Equation (\ref{p}) shows that the volume integration over tilted 
hyperplane (\ref{Ksi}) taken at the observation instant $\tau$ can be 
reduced to the integration over hyperplane $\Sigma_t=\{y\in{\mathbb 
M}_6:y^0=t\}$. Direct calculation shows that the bound six-momentum depends 
on the state of the particle's motion at the end points only:
\begin{eqnarray}
p_{\mbox{\scriptsize bnd}}^\mu &=&\int_{\Sigma_t}
{\rm d}\sigma_0 T_{\mbox{\scriptsize bnd}}^{0\mu}\nonumber\\ 
&=&\frac{e^2}{4\pi^2}
\left[\frac{3}{35}\frac{-12u^0u^\mu +40(u^0)^3u^\mu 
+\eta^{0\mu }\left(-3/2+12(u^0)^2\right)}{(t-u)^3} \right.
\nonumber\\ 
&+&\left.\frac{3}{35}\frac{-5a^\mu + 31a^0u^0u^\mu + 33a^\mu 
(u^0)^2+ 3\eta^{0\mu }a^0}{(t-u)^2}  \right.
\nonumber\\ 
&+&\left.\frac{1}{35}
\frac{37a^0a^\mu + 71a^2u^0u^\mu +\eta^{0\mu 
}a^2}{t-u}\right]_{u\to -\infty}^{u\to t}.
\end{eqnarray}
The lower limit is equal to zero while the upper one tends to infinity.

Now we turn to the calculation over a piecewise surface which consists of 
the cap covering the world tube $\Sigma_r$ and $\Sigma_r$ itself. The cap is 
a part of the observation hyperplane. It is parametrized by the curvilinear 
coordinates (\ref{flat}) where the time parameter $u$ changes from 
$z^0(\tau)$ to $t$ (from the point ${\sf E}$ to the point ${\sf A}$, see 
Fig.\ref{cap}). According to the integration rule (\ref{p}) we pass to MCLF 
where $u^\mu(\tau)=(1,{\bf 0})$ and $a^\mu(\tau)=(0,{\bf a})$ and then 
performs the Lorenz transformation to the laboratory frame.  The expression 
in between the square brackets taken in MCLF and then transformed by 
corresponding Lorentz matrix $\Lambda$ becomes
\begin{equation}
\frac{e^2}{4\pi^2}
\left[
\frac32\frac{u^\mu(\tau)}{(t-z^0(\tau))^3}+
\frac{12}{5}\frac{a^\mu(\tau)}{(t-z^0(\tau))^2} 
+2\frac{a^2u^\mu(\tau)}{(t-z^0(\tau))}
\right].
\end{equation}
This expression coincides with that (\ref{pbnd}) where 
the constant distance $r$ should be replaced by $t-z^0(\tau)$. Indeed, 
the retarded distance $r$ is equal to this value in MCLF (see 
Fig.\ref{cone}). We see that the bound of the cap is sewn with the upper 
bound of the world tube smoothly. 

So far as the upper limit of integration is concerned, we obtain the 
divergent expression where particle's characteristics are evaluated in the 
neighbourhood of point $A$ while the Lorentz matrix - at point $E$ (see 
Fig.\ref{cap}). It is too complicated. The best plan to be followed is to 
take the limit $r\to 0$ in (\ref{pbnd}). It allows us to evaluate the bound 
part of six-momentum in the neighbourhood of the particle without resort to 
the cap covering this thin tube.

The radiative part (\ref{prad}) of energy-momentum carried by the retarded 
electromagnetic field (\ref{F}) does not contain the distance $r$ at all. 
Therefore, the radiation flux across the world tube of an arbitrary radius
$r$ is equal to flux which flows across a space-like surface.

\section{Angular momentum of the retarded Li\'enard-Wi\-e\-chert field in 
six dimensions}
\setcounter{equation}{0} 

We now turn to the calculation of the angular momentum tensor
\begin{equation} \label{M_r}
M_{\mbox{\scriptsize em}}^{\mu\nu} = 
\int_{\Sigma_r}{\rm d}\sigma_\alpha\left(y^\mu T^{\alpha\nu}-y^\nu 
T^{\alpha\mu}\right)\,.
\end{equation}
We calculate how much electromagnetic field angular momentum flows across a 
thin world tube enclosing particle's trajectory (\ref{trj}) up to the 
observation time $\tau$ (see Fig.\ref{tube}).

Decomposition of the angular momentum tensor density into the bound and the 
radiative components is a very important for the calculation. Indeed, the 
former accounts for the angular momentum which remains bound the charge 
while the latter corresponds to the amount of angular 
momentum which escapes to infinity.

We put (\ref{Tr_Tb}) into (\ref{M_r}) where right-hand side of eq.(\ref{y_r})
should be substituted for $y$. It contains the term which is proportional 
to distance $r$. Vector surface element (\ref{sgm}) also depends on $r$. 
In general, terms scaling as $r^2$ may appear. Since $T_{(-4)}^{\alpha\beta}$
is proportional to $k^\alpha k^\beta$ and the equality 
$k_\alpha T_{(-5)}^{\alpha\beta}=0$ is fulfilled, the radiative component
$M_{\mbox{\scriptsize rad}}^{\mu\nu}$ does not depend on the distance:
\begin{eqnarray}
M_{\mbox{\scriptsize 
rad}}^{\mu\nu}&=&\int_{\Sigma_r}{\rm d}\sigma_\alpha\left(z^\mu 
T_{\rm rad}^{\alpha\nu}-z^\nu T_{\rm rad}^{\alpha\mu}\right) 
\nonumber\\
&&+
\int_{\Sigma_r}{\rm d}\sigma_\alpha\left[\left(y^\mu -z^\mu \right)
T_{(-5)}^{\alpha\nu}-\left(y^\nu -z^\nu \right)T_{(-5)}^{\alpha\mu}\right] 
\nonumber\\
&&+\int_{-\infty}^{\tau} {\rm d} u\int {\rm d}\Omega_4
r^6a_kk_\alpha\left(k^\mu T_{(-6)}^{\alpha\nu}-k^\nu T_{(-6)}^{\alpha\mu}
\right)
\nonumber\\
&=&-\int_{-\infty}^{\tau} {\rm d} u\int {\rm d}\Omega_4
r^4u_\alpha\left(z^\mu T_{(-4)}^{\alpha\nu}-z^\nu T_{(-4)}^{\alpha\mu}
\right)
\nonumber\\
&&-\int_{-\infty}^{\tau} {\rm d} u\int {\rm d}\Omega_4
r^5u_\alpha\left(k^\mu T_{(-5)}^{\alpha\nu}-k^\nu T_{(-5)}^{\alpha\mu}
\right)
\nonumber\\
&&+\int_{-\infty}^{\tau} {\rm d} u\int {\rm d}\Omega_4
r^6a_kk_\alpha\left(k^\mu T_{(-6)}^{\alpha\nu}-k^\nu T_{(-6)}^{\alpha\mu}
\right).
\end{eqnarray}
Having performed the angle integration we obtain
\begin{eqnarray} \label{Mrad}
M_{\mbox{\scriptsize rad}}^{\mu\nu}=&&\frac{e^2}{4\pi^2}
\left\{
\int_{-\infty}^{\tau} {\rm d} u
\left(
z^\mu P_{\mbox{\scriptsize rad}}^\nu -
z^\nu P_{\mbox{\scriptsize rad}}^\mu 
\right) 
\right.
\nonumber\\ 
&&\left.+
\int_{-\infty}^{\tau} {\rm d} u
\left[
\frac45\left(a^\mu{\dot a}^\nu-a^\nu{\dot a}^\mu\right)
+\frac{64}{35}a^2\left(u^\mu a^\nu-u^\nu a^\mu\right)\right]\right\}
\end{eqnarray}
where $P_{\mbox{\scriptsize rad}}$ denotes the integrand of 
eq.(\ref{prad}). 

The remaining terms involved in the angular momentum tensor 
density constitute the bound part of the electromagnetic field's
angular momentum:
\begin{eqnarray}
M_{\mbox{\scriptsize 
bnd}}^{\mu\nu}=&&\int_{\Sigma_r}{\rm d}\sigma_\alpha\left(z^\mu 
T_{\rm bnd}^{\alpha\nu}-z^\nu T_{\rm bnd}^{\alpha\mu}\right) 
\nonumber\\
&&+
\int_{\Sigma_r}{\rm d}\sigma_\alpha\left[\left(y^\mu -z^\mu \right)
T_{(-8)}^{\alpha\nu}-\left(y^\nu -z^\nu \right)T_{(-8)}^{\alpha\mu}\right] 
\nonumber\\
&&+
\int_{\Sigma_r}{\rm d}\sigma_\alpha\left[\left(y^\mu -z^\mu \right)
T_{(-7)}^{\alpha\nu}-\left(y^\nu -z^\nu \right)T_{(-7)}^{\alpha\mu}\right] 
\nonumber\\
&&+\int_{-\infty}^{\tau} {\rm d} u\int {\rm d}\Omega_4
r^5(-u_\alpha +k_\alpha)\left(k^\mu T_{(-6)}^{\alpha\nu}-k^\nu 
T_{(-6)}^{\alpha\mu}
\right)
\end{eqnarray}
Volume integration shows that the decomposition is meaningful. Indeed, the 
bound  angular momentum depends on the state of particle's motion at the 
observation instant only:
\begin{equation} \label{Mbnd}
M_{\mbox{\scriptsize bnd}}^{\mu\nu}=\frac{e^2}{4\pi^2}\lim_{r\to 0}\left(
z^\mu P_{\rm bnd}^\nu -z^\nu P_{\rm bnd}^\mu
+\frac{12}{5}\frac{u^\mu a^\nu - u^\nu a^\mu}{r}
\right).
\end{equation}
By symbol $P_{\rm bnd}$ we mean the expression in between the squared 
brackets of eq.(\ref{pbnd}).

It is worth noting that $M_{\mbox{\scriptsize bnd}}^{\mu\nu}$ contains, 
apart from the usual term of type $z\wedge p_{\rm part}$, also an extra term 
which can be interpreted as the "shadow" of internal angular momentum. It 
prompts that the bare "core" possesses a "spin". 

\section{Energy-momentum and angular momentum balance equations}
\setcounter{equation}{0} 
To derive the radiation reaction force in six dimensions we study the 
energy-momentum and angular momentum balance equations.

We calculate how much electromagnetic-field momentum and angular momentum 
flow across hypersurface $\Sigma_r$ up to the proper time $\tau$. We can do 
it at a time $\tau+\triangle\tau$. We demand that change in these quantities 
be balanced by a corresponding change in the particle's ones, so that the 
total energy-momentum and angular momentum are properly conserved.

Expressions (\ref{pbnd}) and (\ref{Mbnd}) show that a charged particle 
cannot be separated from its bound electromagnetic "cloud" which has its own 
energy-momentum and angular momentum. These quantities together with their 
"bare" mechanical counterparts constitute the six-momentum and angular 
momentum of "dressed" charged particle. We proclaim the {\it finite} 
characteristics as those of true physical meaning.

It would make no sense to disrupt the bonds between different powers of 
small parameter $r$ in (\ref{pbnd}). It is sufficient to assume that a 
charged particle possesses its own (already renormalized) six-momentum 
$p_{\rm part}$ which is transformed as an usual six-vector under the 
Poincar\'e group. The total energy-momentum of a closed system of an 
arbitrarily moving charge and its electromagnetic field is equal to the sum 
\begin{equation} \label{Pmn}
P^\mu =p_{\rm part}^\mu +p_{\rm rad}^\mu 
\end{equation} 
where $p_{\rm rad}$ is the radiative part (\ref{prad}) of electromagnetic 
field's energy-momentum which detaches itself from the charge and leads an 
independent existence. 

With (\ref{Mbnd}) in mind we assume that already renormalized angular 
momentum tensor of the particle has the form 
\begin{equation} \label{Mprt}
M_{\rm part}^{\mu\nu} =z^\mu p_{\rm part}^\nu-z^\nu p_{\rm part}^\mu +
u^\mu\pi_{\rm part}^\nu -u^\nu\pi_{\rm part}^\mu .
\end{equation} 
In \cite{Pl,Pv,NFS} the extra momentum $\pi_{\rm part}$ is due to additional 
degrees of freedom associated with acceleration involved in Lagrangian 
function for rigid particle.

Total angular momentum of our composite particle plus field system
is written as
\begin{equation} \label{Mmn}
M^{\mu\nu}=M_{\rm part}^{\mu\nu} + M_{\mbox{\scriptsize rad}}^{\mu\nu}
\end{equation} 
where $M_{\rm rad}$ is the radiative part (\ref{Mrad}) of electromagnetic 
field's angular momentum which depends on all previous motion of a source.
Our next task is to derive expressions which explains how six-mo\-men\-tum
and angular momentum of charged particle depend on its velocity and 
acceleration etc. Via the differentiation of eq.(\ref{Pmn}) we obtain
the following energy-momentum balance equation:
\begin{equation} \label{pdot}
{\dot p}_{\rm part}^\mu = -\frac{e^2}{4\pi^2}
\left(\frac45u^\mu\dot{a}^2-\frac{6}{35}a^2\dot{a}^\mu
+\frac37a^\mu (a^2)^\cdot+2a^4u^\mu\right).
\end{equation} 
(All the particle characteristics are evaluated at the instant of 
observation $\tau$.) Having differentiated (\ref{Mmn}) and taking into 
account (\ref{pdot}) we arrive at the equality which does not contain ${\dot 
p}_{\rm part}$:
\begin{eqnarray} \label{Mdot}
&&u^\mu\left(p_{\rm part}^\nu +{\dot \pi}_{\rm part}^\nu\right)
-u^\nu\left(p_{\rm part}^\mu +{\dot \pi}_{\rm part}^\mu\right)
+a^\mu\pi_{\rm part}^\nu - a^\nu\pi_{\rm part}^\mu
\nonumber\\ 
&=&-\frac{e^2}{4\pi^2}
\left[\frac45\left(a^\mu\dot{a}^\nu 
-a^\nu\dot{a}^\mu\right)+\frac{64}{35}a^2\left(u^\mu a^\nu -u^\nu a^\mu 
\right)\right].
\end{eqnarray} 
The rank of this system of fifteen linear equations in twelve variables
$p_{\rm part}^\alpha +{\dot \pi}_{\rm part}^\alpha$ and $\pi_{\rm 
part}^\alpha$ is equal to nine. It is convenient to rewrite them as follows
\begin{equation}
u\wedge (p+\dot\pi) +a\wedge\pi=-\frac{e^2}{4\pi^2}
\left[\frac45 a\wedge\dot a  +\frac{64}{35}a^2 u\wedge a\right]
\end{equation}
where symbol $\wedge$ denotes the wedge product. Hence one has again
\begin{equation}
u\wedge \left(p+\dot\pi +\frac{e^2}{4\pi^2}\frac{64}{35}a^2 a\right) 
+a\wedge\left(\pi+\frac{e^2}{4\pi^2}\frac45\dot a\right) =0.
\end{equation}
Their solutions involve three arbitrary scalar functions, say $M$, $\mu$ and
$\nu$:
\begin{eqnarray} \label{pp}
 p_{\rm part}^\beta +{\dot \pi}_{\rm part}^\beta =Mu^\beta +\nu a^\beta -
\frac{e^2}{4\pi^2}\frac{64}{35}a^2 a^\beta
\\ 
 \pi_{\rm part}^\beta =\mu a^\beta +\nu u^\beta -
\frac{e^2}{4\pi^2}\frac45{\dot a}^\beta . \label{pipi}
\end{eqnarray} 

Scrupulous analysis of their consistency with six first-order 
differential equations (\ref{pdot}) reveals, that six-momentum of charged 
particle contains two (already renormalized) constants (see Appendix A):
\begin{equation} \label{p_ost}
p_{\rm part}^\beta = mu^\beta +\mu\left(-{\dot a}^\beta 
+\frac32a^2u^\beta\right)
+\frac{e^2}{4\pi^2}\left[\frac45{\ddot a}^\beta
-\frac85u^\beta (a^2)^\cdot-\frac{64}{35}a^2a^\beta\right].
\end{equation}
The first, $m$, looks as a rest mass of the charge. But the true rest 
mass is identical to the scalar product of the six-momentum and 
six-velocity \cite{KosPr}. Since the scalar product depends on 
the square of acceleration as well as its time derivative
\begin{equation} 
m_0=-(p_{\rm part}\cdot 
u)=m+\frac{\mu}{2}a^2-\frac{e^2}{4\pi^2}\frac25(a^2)^\cdot 
\end{equation}
the renormalization constant $m$ is formal parameter and its 
physical sense is not clear.

The second, $\mu$, is intimately connected with the wedge product 
$u\wedge\pi_{\rm part}:=s_{\rm part}$. With eq.(\ref{Mprt}) in mind 
we call 
\begin{equation} \label{ss}
s_{\rm part}^{\alpha\beta} = \mu\left(u^\alpha a^\beta -u^\beta 
a^\alpha\right) -\frac{e^2}{4\pi^2}\frac45\left(u^\alpha{\dot a}^\beta
-u^\beta{\dot a}^\alpha \right)
\end{equation} 
the internal angular momentum  of the particle.
But its magnitude is not constant:
\begin{equation} \label{s^2}
s^2=-\frac12s^{\rm part}_{\alpha\beta}s_{\rm part}^{\alpha\beta}
=\mu^2a^2+\mu\frac{e^2}{5\pi^2}(a^2)^\cdot +\frac{e^4}{25\pi^4}\left(
\dot{a}^2+a^4\right).
\end{equation}
Therefore, this name can not be understand literally.

Having substituted the right-hand side of eq.(\ref{p_ost}) for the 
particle's six-momentum in eq.(\ref{pdot}) we derive the Lorentz-Dirac 
equation of motion of a charged particle under the influence of its own 
electromagnetic field. The problem of including of an external device 
requires careful consideration.

When considering the system under the influence of an external device the 
time derivative $\dot{P}$ of total momentum $P$ is equal to external 
force $F_{ext}$. It changes the energy-momentum balance equation 
(\ref{pdot}) as follows:
\begin{equation} \label{Fext}
{\dot p}_{\rm part}^\mu +\frac{e^2}{4\pi^2}
\left(\frac45u^\mu\dot{a}^2-\frac{6}{35}a^2\dot{a}^\mu
+\frac37a^\mu (a^2)^\cdot+2a^4u^\mu\right)=F_{ext}^\mu .
\end{equation} 
Corresponding change of the total angular momentum $M^{\mu\nu}$ is defined 
by an external torque:
\begin{equation} \label{Mext}
\dot{M}^{\mu\nu}=z^\mu F_{ext}^\nu - z^\nu F_{ext}^\mu .
\end{equation}

Expression (\ref{p_ost}) was firstly obtained by Kosyakov in 
\cite[eq.(37)]{Kos}. The derivation is based upon consideration of 
energy-momentum conservation only. The author constructs an appropriate 
Schott term to ensure the orthogonality of the radiation reaction force to 
the particle six-velocity.

\section{Conclusions}
\setcounter{equation}{0} 
Our consideration is founded on the field and the interaction terms of the 
action (\ref{I}). They constitute the action functional which governs the 
propagation of the electromagnetic field produced by a moving charge 
(i.e. the Maxwell equations with point-like source).

A surprising feature of the study of Poincar\'e-invariance of the dynamics 
of a closed particle plus field system is that the conservation laws 
determines the form of particle's individual characteristics such as the 
momentum and the angular momentum. The matter is that a charged particle 
cannot be separated from its bound electromagnetic "cloud" which has its own 
momentum and angular momentum. These quantities together with corresponding 
characteristics of bare "core" constitute the momentum and angular momentum 
of "dressed" charged particle.

So, in four dimensions the momentum of a bare "core" is proportional 
to its four-velocity. An electromagnetic "cloud" renormalizes the bare 
mass and adds the term which is proportional to the four-acceleration 
(see eq.(\ref{Texp})). The extra term can be obtained from the angular 
momentum balance equation \cite{Yar}. In six dimensions a bare charge should 
possess (non-conventional) internal angular momentum 
\begin{equation}\label{s6}
s_0^{\mu\nu} = 
\mu_0\left(u^\mu a^\nu-u^\nu a^\mu\right) 
\end{equation}
with magnitude which is proportional to the square of acceleration (see 
eq.(\ref{s^2})).
Its six-momentum is not proportional to six-velocity:
\begin{equation}\label{bm6}
p_0^\mu = m_0u^\mu + 
\mu_0\left(-{\dot a}^\mu+\frac32a^2u^\mu\right).
\end{equation}
The energy-momentum and angular momentum balance equations give the 
six-momentum (\ref{p_ost}) of "dressed" charged particle which coincides 
with that obtained in \cite{Kos}.

We see that the particle part of initial action integral (\ref{I}) 
which is proportional to the worldline lenght is inconsistent with the 
others $I_{\mbox{\scriptsize field}}$ and $I_{\mbox{\scriptsize int}}$. 
The action functional based on the higher-derivative Lagrangian for a 
"rigid" relativistic particle \cite{Pl,Pv,NFS,Nest} should be 
substituted for $I_{\mbox{\scriptsize particle}}$ in (\ref{Mp}). It is 
sufficient to renormalize {\em all} the divergences arising in 
six-dimensional electrodynamics (these connected with bound six-momentum 
and those associated with bound angular momentum of the electromagnetic 
field).
The variation of modified action with respect to particle variables results 
the appropriate equation of motion of a charged particle in response to the 
electromagnetic field.

It is interesting to consider the motion of test particles (i.e. point 
charges which themselves do not influence the field). In four dimensions the 
limit $e\to 0$ and $m\to 0$ with their ratio being fixed results the 
Maxwell-Lorentz theory of test particles. The momentum of test particle is 
proportional to its four-velocity, the loss of energy due to radiation is 
too small to be observed. In six dimensions the test particle is the {\em 
rigid} particle. Its momentum is not parallel to six-velocity. The problem 
of motion of such particles in an external electromagnetic field is 
considered in \cite{Nst}.

The results of the paper can be extended in an arbitrary even dimensions.
We can limit our calculations to the radiative parts of energy-momentum and 
angular momentum carried by electromagnetic field. We are sure that the 
energy-momentum and angular momentum balance equations allow us to 
establish the radiation reaction force. 

\section*{Acknowledgments}

The author would like to thank B.P.Kosyakov,
V.Tre\-tyak,  and Dr.A.Du\-vi\-ryak for 
helpful discussions and critical comments.

\appendix
\renewcommand{\thesubsection}{Appendix \Alph{subsubsection}}
\renewcommand{\theequation}{\Alph{subsection}.\arabic{equation}}  
\subsection{}
\setcounter{equation}{0} 
Since $(u\cdot a)=0$, the scalar product of particle six-velocity on the 
first-order time derivative of particle six-momentum (\ref{pdot}) is as
follows:
\begin{equation} \label{A1}
({\dot p}_{\rm part}\cdot u) = \frac{e^2}{4\pi^2}
\left(\frac45\dot{a}^2+\frac{64}{35}a^4\right).
\end{equation} 
Similarly, the scalar product of particle acceleration on the 
particle six-momentum (\ref{pp}) is given by 
\begin{equation} \label{A2}
(p_{\rm part}\cdot a) = \nu a^2 -
\frac{e^2}{4\pi^2}\frac{64}{35}a^4 - ({\dot \pi}_{\rm part}\cdot a)
\end{equation}
Summing up (\ref{A1}) and (\ref{A2}) we obtain 
\begin{equation} \label{A3}
(p_{\rm part}\cdot u)^\cdot = \nu a^2 + \frac{e^2}{4\pi^2}
\frac45\dot{a}^2 - ({\dot \pi}_{\rm part}\cdot a).
\end{equation}
From the other hand, the time derivative $(p_{\rm part}\cdot u)^\cdot$ of 
scalar product of particle velocity on the momentum (\ref{pp}) is written as
\begin{equation} \label{A4}
(p_{\rm part}\cdot u)^\cdot = -{\dot M} - ({\dot \pi}_{\rm part}\cdot 
u)^\cdot .
\end{equation}
Subtracting (\ref{A4}) from (\ref{A3}), one has again
\begin{equation} \label{A5}
{\dot M} = -\nu a^2 - \frac{e^2}{4\pi^2}
\frac45\dot{a}^2- ({\ddot \pi}_{\rm part}\cdot u).
\end{equation}

Further we calculate the scalar product of the second-order derivative of
(\ref{pipi}) on particle's velocity:
\begin{equation} \label{A6}
({\ddot \pi}_{\rm part}\cdot u) =-2\dot\mu a^2 -\frac32\mu(a^2)^\cdot
-\ddot\nu -\nu a^2 +
\frac{e^2}{4\pi^2}\left(\frac85(a^2)^{\cdot\cdot}-\frac45\dot a^2\right) .
\end{equation} 
Having substituted it into previous equation we arrive at the following 
differential equation:
\begin{equation} \label{A7}
{\dot M} = 2\dot\mu a^2 +\frac32\mu (a^2{\dot )} +\ddot\nu  - 
\frac{e^2}{4\pi^2}\frac85(a^2)^{\cdot\cdot} .
\end{equation}
It can be solved iff the scalar $\mu$ does not change with time:
\begin{equation} \label{A8}
M = m + \frac32\mu a^2 +\dot\nu  - 
\frac{e^2}{4\pi^2}\frac85(a^2)^\cdot .
\end{equation}

Having substituted it into (\ref{pp}) and taking into account the time 
derivative of (\ref{pipi}), we derive the expression (\ref{p_ost}) for 
the components of six-momentum of charged particle. It depends on two 
renormalization constants, $m$ and $\mu$.

\end{document}